\documentclass{article} 
\usepackage{iclr2027_conference,times}
\usepackage{booktabs}
\usepackage{caption}
\usepackage{multirow}
\usepackage{listings}
\usepackage[most]{tcolorbox}
\usepackage[utf8]{inputenc} 
\usepackage[T1]{fontenc}    
\usepackage{hyperref}       
\usepackage{url}            
\usepackage{booktabs}       
\usepackage{amsfonts}       
\usepackage{nicefrac}       
\usepackage{microtype}      
\usepackage{xcolor}         
\usepackage{colortbl}
\usepackage{pifont}
\usepackage{multirow}
\definecolor{darkgreen}{rgb}{0, 0.5, 0} 
\definecolor{darkred}{rgb}{0.8, 0, 0}  
\newcommand{\greencheck}{\textcolor{darkgreen}{\ding{51}}}

\usepackage{xspace}
\usepackage{fontawesome}
\usepackage{ulem}
\usepackage{enumitem}

\definecolor{darkblue}{rgb}{0, 0, 0.5}
\definecolor{veronica-red}{RGB}{196,30,58}
\hypersetup{colorlinks=true, citecolor=darkblue, linkcolor=darkblue, urlcolor=darkblue}
\definecolor{ForestGreen}{RGB}{34,139,34}
\definecolor{BrickRed}{rgb}{.72,0,0}
\definecolor{LakeBlue}{RGB}{0,61,153}
\definecolor{boxgreen}{rgb}{0.9,1.0,0.9}
\definecolor{boxblue}{rgb}{0.9,0.9,1.0}
\definecolor{boxred}{rgb}{1.0, 0.85, 0.95}

\usepackage{amsmath,amsfonts,bm}

\def\eqref#1{equation~\ref{#1}}

\def\1{\bm{1}}

\DeclareMathAlphabet{\mathsfit}{\encodingdefault}{\sfdefault}{m}{sl}
\SetMathAlphabet{\mathsfit}{bold}{\encodingdefault}{\sfdefault}{bx}{n}

\usepackage{fvextra}
\usepackage[most]{tcolorbox}
\newcommand{\promptfile}[3][]{%
  \begin{tcolorbox}[
    title={#2},
    breakable,
    enhanced,
    colback=gray!4,
    colframe=gray!65!black,
    colbacktitle=gray!65!black,
    fonttitle=\bfseries\small,
    boxrule=0.6pt,
    arc=2pt,
    left=4pt, right=4pt, top=2pt, bottom=2pt,
    #1]
  \VerbatimInput[
    fontsize=\scriptsize,
    breaklines=true,
    breakanywhere=true,
    breakindent=0pt,
    breaksymbolleft={\textcolor{gray}{\tiny$\hookrightarrow$}},
  ]{#3}
  \end{tcolorbox}%
}

\usepackage{hyperref}
\usepackage{url}
\usepackage{graphicx}
\usepackage{float}
\title{OpenMobile-2: Building Versatile Mobile Agents with Scalable Environments and App-Native Tools}

\author{%
\textbf{Chenyang Yan}$^{12\,*}$\:
\textbf{Yingying Zhang}$^{2\,*}$\:
\textbf{Kanzhi Cheng}$^{12\,*\,\dagger}$\:
\textbf{Qiushi Sun}$^{3\,*}$\:
\textbf{Zhengyuan Pan}$^{2\,*}$\\
\textbf{Zheng Ma}$^{2\,*\,\dagger}$
\textbf{Hang Yan}$^{4}$
\textbf{Yian Wang}$^{5}$
\textbf{Nuo Chen}$^{1}$
\textbf{Jialin Cao}$^{1}$
\textbf{Xingdong Gong}$^{1}$
\textbf{Wenpo Song}$^{1}$\\
\textbf{Han Chen}$^{1}$\:
\textbf{Zichen Ding}$^{6}$\:
\textbf{Fangzhi Xu}$^{4}$\:
\textbf{Shujian Huang}$^{1}$\:
\textbf{Xinyu Dai}$^{1}$\:
\textbf{Yichen Liu}$^{2}$\\
\textbf{Tiankuo Yao}$^{2}$\:
\textbf{Bo Wang}$^{2}$\:
\textbf{Ben Kao}$^{3}$\:
\textbf{Jianbing Zhang}$^{1\,\ddagger}$\:
\textbf{Lewei Lu}$^{2\,\ddagger}$\:
\textbf{Dahua Lin}$^{2}$\\[0.8em]
$^{1}$Nanjing University\quad
$^{2}$SenseTime\quad
$^{3}$The University of Hong Kong\quad
$^{4}$Xi'an Jiaotong University\\
$^{5}$National University of Singapore\quad
$^{6}$East China Normal University
}

\makeatletter
\newcommand{\authornotes}[1]{\begingroup
  \renewcommand{\thefootnote}{}%
  \renewcommand{\@makefntext}[1]{\noindent##1}%
  \footnotetext{#1}\endgroup}
\makeatother

\newsavebox{\fighybridbox}

\iclrfinalcopy
\ClearShipoutPicture
\begin{document}

\setlength{\tabcolsep}{0pt}%
\maketitle
\vspace{-12pt}
\setlength{\tabcolsep}{6pt}
\authornotes{\textsuperscript{*}Equal contribution\quad 
\textsuperscript{\dag}Proj lead\quad
\textsuperscript{\ddag}Corresponding author}
\fancyhf{}
\lhead{Preprint. Under review.}
\renewcommand{\headrulewidth}{0.4pt}
\fancyfoot[C]{\thepage}

\begin{abstract}

We introduce OpenMobile-2, a near-frontier mobile agent with fully open training environments and recipes. We make three key advances: (1) \textit{Diverse environments with simulated commercial apps}: We build \textsc{MobileGym++}, featuring 35 realistic, functionally rich commercial-style apps with cross-app workflows, while preserving full controllability for reset and verification. Together with newly configured apps in Android emulators, this yields a diverse playground spanning over 110 apps. (2) \textit{Open training data at scale}: Building on this foundation, we curate nearly 12K mobile interaction trajectories for supervised fine-tuning and the largest open collection of verifiable RL training data for mobile agents, comprising over 2K executable tasks with automatic rewards. (3) \textit{Hybrid GUI and app-native tool use}: We explore an experimental mobile-use setting where apps expose selected functionalities as app-native tools alongside their GUIs, allowing agents to interleave GUI actions and tool calls within a task. We implement this setting in \textsc{MobileGym++} with over $300$ carefully scoped tools across $50+$  apps. Additionally, we introduce \textsc{MobileGym++ Bench} for complex, long-horizon commercial mobile scenarios, supporting both GUI-only and hybrid GUI–tool evaluation on a shared task suite. 
OpenMobile-2 performs competitively across established benchmarks, including AndroidWorld (79.9) and MobileWorld (50.4), while showing promising transfer to real-device mobile use,
nearly doubling SPA-Bench performance from 31.9 to 59.6.
All resources are available at \href{https://os-copilot.github.io/OpenMobile2-Home/}{OpenMobile-2}.

\end{abstract}

{\setlength{\intextsep}{6pt}
\begin{figure}[h]
    \centering
    \includegraphics[
        width=\linewidth
    ]{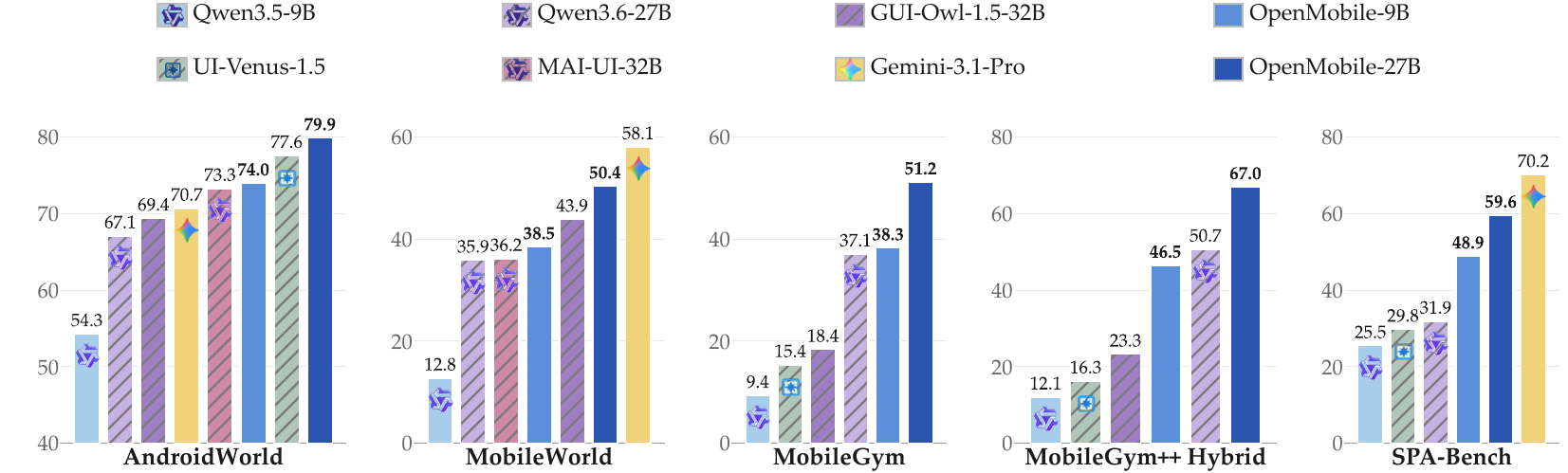}
    \caption{OpenMobile-2 performs strongly across established mobile agent benchmarks and achieves leading results among open-source models on real-device evaluation.}
    \label{fig:intro-overview}
\end{figure}}

\begin{figure}[t]
\centering

\begin{minipage}[b]{0.38\linewidth}
    \centering
    \includegraphics[width=\linewidth]{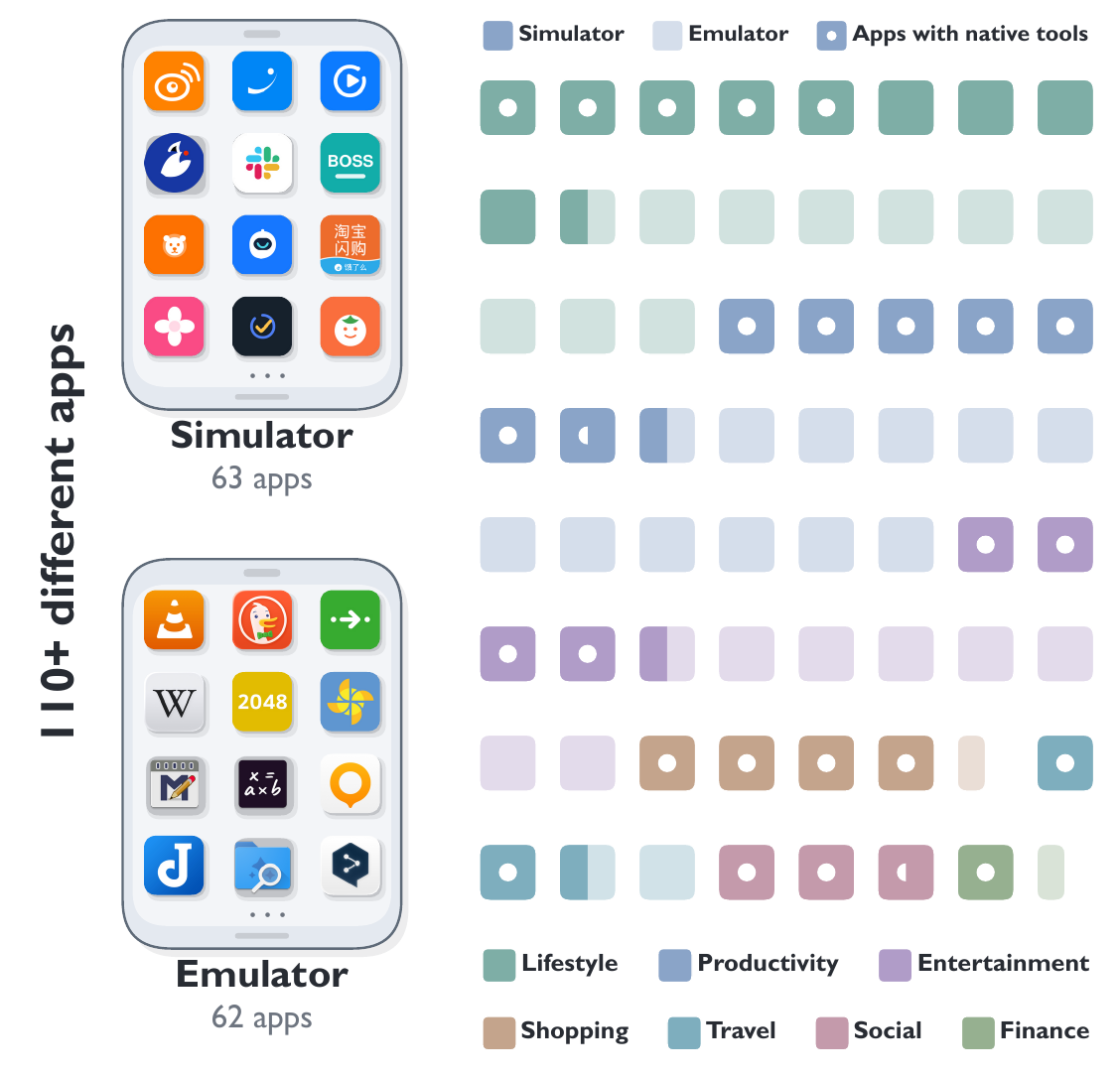}
    \\[0.3em]
    {\small (a) Application coverage}
\end{minipage}
\hfill
\begin{minipage}[b]{0.56\linewidth}
    \centering
    \includegraphics[width=\linewidth]{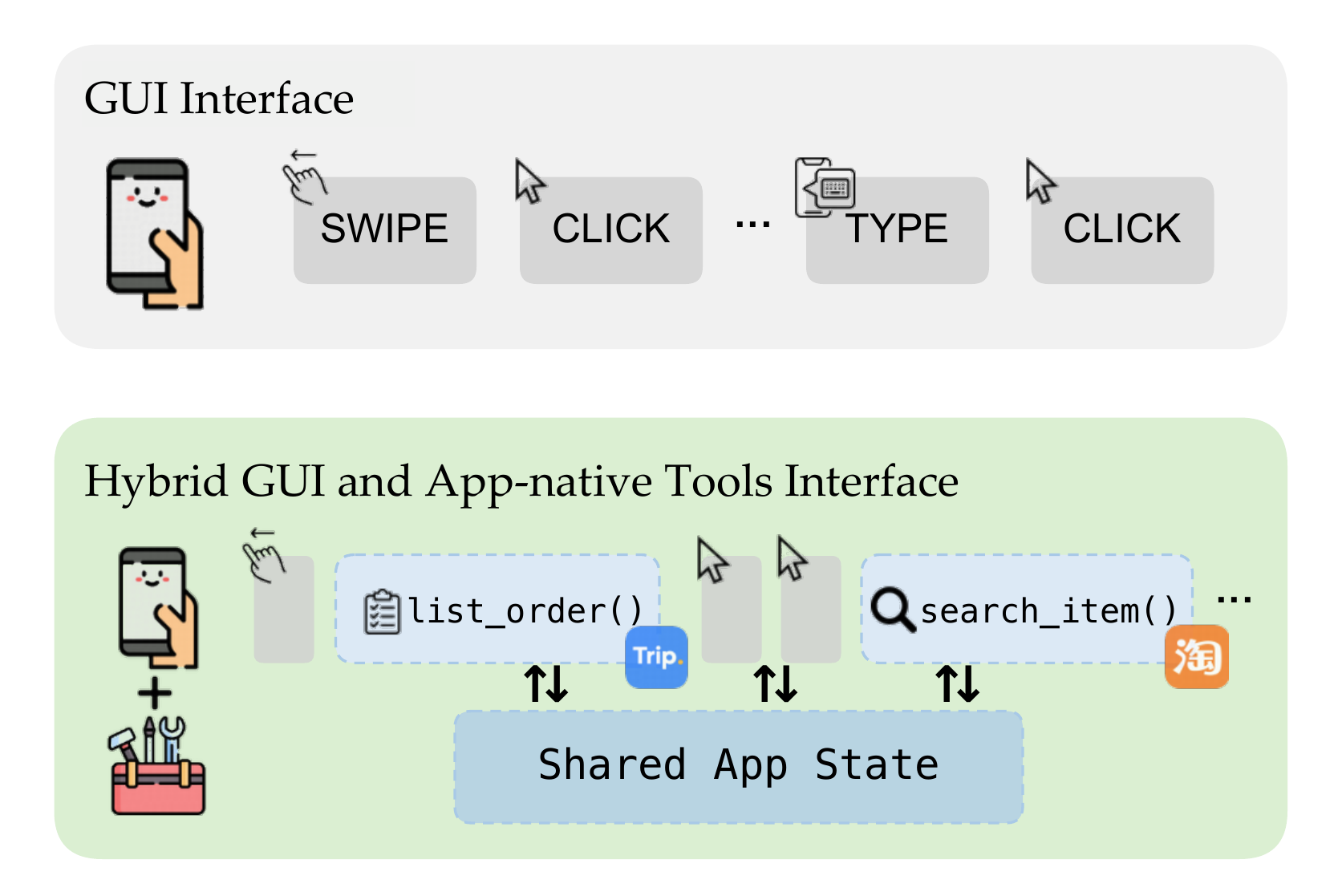}
    \\[0.3em]
    {\small (b) GUI-only vs.\ Hybrid GUI and app-native tools}
\end{minipage}

\caption{(a)~Application coverage of OpenMobile-2. Each square represents two applications; dark and light tiles denote simulator and emulator apps, and dots mark apps with native tools.
(b)~Hybrid GUI and app-native tool use. The same task can be completed through GUI-only interaction (top) or by interleaving GUI actions with app-native tool calls (bottom) over a shared app state.}
\label{fig:intro-hybrid}
\end{figure}

\section{Introduction}
\label{sec:intro}

Mobile agents built on vision-language models are increasingly capable of understanding smartphone interfaces and automating everyday tasks~\citep{wang2025ui, zhou2025mai, xu2026mobile, gao2026ui, zhou2026qwen}.
Such progress depends heavily on large-scale interaction experience, typically in the form of trajectories collected across diverse tasks, applications, and environments~\citep{cheng2026openmobile, team2026ui}. 
Gathering such experience independently is costly, while high-quality interaction data remains largely proprietary, leaving the open community with limited training resources and open models trailing well behind.

Scaling interaction experience without costly human annotation requires a diverse and controllable mobile playground. 
Especially for commercial applications, reproducing rich workflows and persistent user states at scale has remained a key obstacle.
Existing open efforts~\citep{sun2025genesis,cheng2026openmobile} largely rely on a small set of applications in Android emulators, mostly open-source apps and system utilities. Such environments provide realistic interaction, but third-party apps often lack reliable reset, while the full Android stack is costly to replicate for large-scale rollout and reinforcement learning~\citep{rawles2025androidworld}. Recent simulated environments offer an alternative path~\citep{wu2026mobilegym, liu2026phoneworld}, providing efficient reset and parallelization, yet their applications remain few and shallow, with no scalable recipe for extending them.

We introduce OpenMobile-2, an open stack of environments, data, and training recipes that puts environment scaling into practice, as shown in Figure~\ref{fig:intro-hybrid}.
We build \textsc{MobileGym++}, a large-scale mobile environment featuring diverse, high-fidelity commercial-style applications, built upon the simulation framework of MobileGym~\citep{wu2026mobilegym}.
In particular, we add 35 functional re-implementations of commercial apps with rich internal state and cross-app workflows, while preserving full controllability for clean reset and verification of task outcomes.
Together with newly configured apps in Android emulator, this forms a diverse training playground spanning over 110 apps. Building on this playground, we curate nearly 12K trajectories for supervised fine-tuning and more than 2K executable tasks for reinforcement learning, each equipped with a reliable verifier.

We further explore an experimental mobile-use setting that combines GUI interaction with app-native tool use. This setting reflects an emerging shift in computer use, where universal but inefficient GUI interaction is increasingly complemented by efficient yet functionally bounded structured interfaces~\citep{hu2026toolcua, shi2026cua}. In \textsc{MobileGym++}, we expose selected app functionalities through over $300$ app-native tools across $50+$ apps, e.g., \textit{search\_course} in a fitness app. This enables us to train agents that interleave GUI actions and tool calls within a task.

To push mobile agent evaluation toward real-world commercial workflows and hybrid GUI--tool interaction, we introduce \textsc{MobileGym++ Bench}, a suite of 215 long-horizon cross-app tasks with persistent user states and programmatic verification. Every task can be solved with or without tools, so GUI-only and hybrid agents are compared on the same tasks.
OpenMobile-2 achieves competitive results on established mobile agent benchmarks, reaching 79.9 on AndroidWorld and 50.4 on MobileWorld.
It further transfers to real-device settings, improving SPA-Bench performance from 31.9 to 59.6 and outperforming strong open-source baselines.
Further analysis highlights the importance of scaling application diversity for improving agent generalization and reveals the complementary benefits of hybrid GUI and app-native tool use.

Our main contributions are:
\begin{itemize}[leftmargin=2em, itemsep=2pt, topsep=0pt, parsep=0pt]

\item \textbf{Environment Scaling with Replicated Applications.}  MobileGym++ faithfully replicates 35 commercial applications with accessible state in an extendable 110-application playground, overcoming the limited scope of existing open environments.

\item \textbf{Open Training Data and Recipes at Scale.}
We release a fresh collection of nearly 12K diverse trajectories and full recipes, together with over 2K executable RL tasks with reliable verifiers, which largely advance training mobile agents.

\item \textbf{A Unified Framework for GUI and App-native Tool Use.} We integrate over $300$ scoped app-native tools across $50+$ apps and introduce MobileGym++ Bench, where GUI-only and hybrid GUI-tool agents can be trained and evaluated on the same long-horizon tasks. 

\end{itemize}

\section{Related Work}
\label{sec:related}

\paragraph{Mobile Agents.} 
Mobile agents follow natural language instructions to operate smartphones through actions such as taps, swipes, and typing.
Early systems rely on proprietary multimodal models orchestrated by external pipelines for screen parsing, planning, and memory~\citep{wang2024mobile, zhang2025appagent, yang2026symphony, sun2026sentinel}, requiring no task specific training.
Native agents instead pretrain vision language models for GUI grounding~\citep{cheng2024seeclick, hong2024cogagent, wu2025atlas} and fine-tune them into end-to-end action models~\citep{qin2025ui, liu2024autoglm, ye2025mobile}.
More recent systems further improve these models through multi-turn reinforcement learning on live devices and interactive sandboxes~\citep{wang2025ui, zhou2025mai, team2026ui, xu2026mobile, gao2026ui}.
This progression makes interaction experience a central resource for scaling mobile agents, spanning high-quality trajectories for supervised training, efficiently resettable environments, and verifiable tasks for reinforcement learning.

\paragraph{Trajectory Data and Interactive Environments.}

Existing training trajectories are primarily obtained in two ways. Human-recorded trajectories~\citep{rawles2023androidinthewild, li2024effects, zhang2024android, lu2025guiodyssey} are fixed once collected, expensive to annotate, and carry no executable notion of success. Interactive environments supply the missing loop and differ in how their apps are obtained. One family runs real apps on Android emulators or physical devices and grades tasks against device state~\citep{rawles2025androidworld, xu2025androidlab, chai2026a3, chen2024spa}, which is realistic but resists clean reset and is costly to parallelize; benchmarks over real commercial apps~\citep{wu2026mobilebench, feng2026mobiflow, sui2026androiddaily} inherit these constraints. The other rebuilds apps as controllable mock clients~\citep{trivedi2024appworld, wu2026mobilegym, liu2026phoneworld}, trading fidelity for cheap parallelism, clean reset, and direct state checks. Both are now used as data factories, synthesizing trajectories and verifiable tasks from the environment itself~\citep{sun2025genesis, cheng2026openmobile, liu2026scalewob, cao2026gui, team2026hymobileagent}, making application coverage a key bottleneck for scaling interaction experience.
Yet applications in existing environments remain shallow, with no scalable way to capture interaction experience across commercial workflows.

\paragraph{Hybrid GUI and Tool Use.} 

GUI actions reach any visible control but are slow and brittle; structured calls are efficient but cover only the functions an interface exposes.
Structured interfaces such as CLIs and MCP are increasingly available alongside GUIs, and recent computer-using agents have begun to combine the two~\citep{sun2026scienceboard, hu2026toolcua, yan2025mcpworld, jia2026osworld, fan2026screenshots}.
On mobile the same split appears as API-only worlds with no phone GUI~\citep{trivedi2024appworld}, as external MCP servers attached to a GUI agent~\citep{kong2026mobileworld}, as shortcuts such as RPA over real apps~\citep{zhao2026mas}, or as actions an agent generates from its own GUI traces~\citep{lee2026improving}.
Prior hybrid approaches typically attach external tools or system-level interfaces to a GUI agent, rather than functions exposed by the app itself. In contrast, MobileGym++ simulates commercial apps that selectively expose their own app-native tools alongside the GUI, enabling tool calls and GUI actions to interleave within a task over shared application state.

\section{\textsc{MobileGym++} Playground}
\label{sec:mobilegym++}
\label{sec:method}

\subsection{A gap in controllable commercial applications}
\label{sec:mgpp-env}

Mobile agents are increasingly moving toward real-world workflows across commercial applications, such as travel booking on Trip and online shopping on Taobao, where tasks are often long-horizon and grounded in real user needs. However, training and evaluation environments for such commercial applications remain largely missing. 
Real-device or Android-emulator setups introduce login friction, irreversible side effects, and costly reset procedures that make large-scale training impractical~\citep{rawles2025androidworld,kong2026mobileworld}. Prior simulated-app work, meanwhile, covers only a narrow range of commercial domains~\citep{wu2026mobilegym,trivedi2024appworld}.
As a result, mobile agents have limited opportunities to practice and be evaluated on realistic mobile-use scenarios.

We built MobileGym++, a browser-hosted mobile environment that supports realistic commercial workflows in controllable simulated applications. The runtime extends MobileGym~\citep{wu2026mobilegym}, preserving its advantages in reset and programmatic evaluation while substantially expanding domain coverage, the number and richness of applications, and support for native cross-app interaction. Observations consist of raw screen pixels; there is no accessibility tree and the agent cannot read business state directly. Device state is fully serializable, enabling deterministic reset and lightweight parallel instantiation. We newly built 35 high-fidelity commercial applications in this round, bringing the total to 63 installable clients spanning travel, shopping, finance, content, social, office, and local services. Each application maintains persistent business state---orders, budgets, itineraries---that evolves through interaction and supports programmatic task verification.
Table~\ref{tab:mgpp-stack-compare} compares MobileGym++ with existing mobile agent environments.

\begin{table*}[t]
\caption{Comparison of mobile agent environments across scale, interaction, and controllability. \ding{51} denotes supported, \ding{55} unsupported, and $\sim$ indicates limited system level cross-app interaction support in MobileGym. \# Comm. Apps and \# Tasks denote the number of covered commercial apps and benchmark tasks, respectively;
Detailed application statistics are provided in Appendix~\ref{app:mgpp-apps}. 
MobileGym++ provides diverse, high-fidelity commercial-style applications with full controllability, while introducing an experimental app-native tool interface.}
\label{tab:mgpp-stack-compare}
\centering
\scriptsize
\setlength{\tabcolsep}{3.0pt}
\renewcommand{\arraystretch}{1.16}
\resizebox{\linewidth}{!}{%
\begin{tabular}{@{}l l c c c c c c@{}}
\toprule
\multirow{2}{*}[-0.6ex]{Environment} 
& \multirow{2}{*}[-0.6ex]{Platform}
& \multicolumn{2}{c}{Scale} 
& \multicolumn{3}{c}{Interaction} 
& \multirow{2}{*}[-0.6ex]{\shortstack{Full\\Controllability}} \\
\cmidrule(lr){3-4} 
\cmidrule(lr){5-7}
& & \# Comm. Apps & \# Tasks  
& Cross-app & GUI & App-native Tools 
& \\
\midrule
AppWorld~\citep{trivedi2024appworld} & API & 9 & 750  
& \ding{51} & \ding{55} & \ding{55} & \ding{51} \\

AndroidLab~\citep{xu2025androidlab} & Emulator & 5 & 138  
& \ding{55} & \ding{51} & \ding{55} & \ding{55} \\

AndroidWorld~\citep{rawles2025androidworld} & Emulator & 0 & 116  
& \ding{51} & \ding{51} & \ding{55} & \ding{55} \\

SPA-Bench~\citep{chen2024spa} & Real-device & 58 & 340  
& \ding{51} & \ding{51} & \ding{55} & \ding{55} \\

MobileWorld~\citep{kong2026mobileworld} & Emulator & 4 & 201  
& \ding{51} & \ding{51} & \ding{55} & \ding{55} \\

MobileGym~\citep{wu2026mobilegym} & Simulator & 12 & 256  
& $\sim$ & \ding{51} & \ding{55} & \ding{51} \\

\textbf{MobileGym++ (ours)} & Simulator & \textbf{35} & \textbf{215}  
& \greencheck & \greencheck & \greencheck & \greencheck \\

\bottomrule
\end{tabular}%
}
\vspace{2pt}
\end{table*}

\subsection{35 high-fidelity commercial applications}
\label{sec:mgpp-apps}

\begin{figure}[t]
\centering
\includegraphics[width=0.85\linewidth]{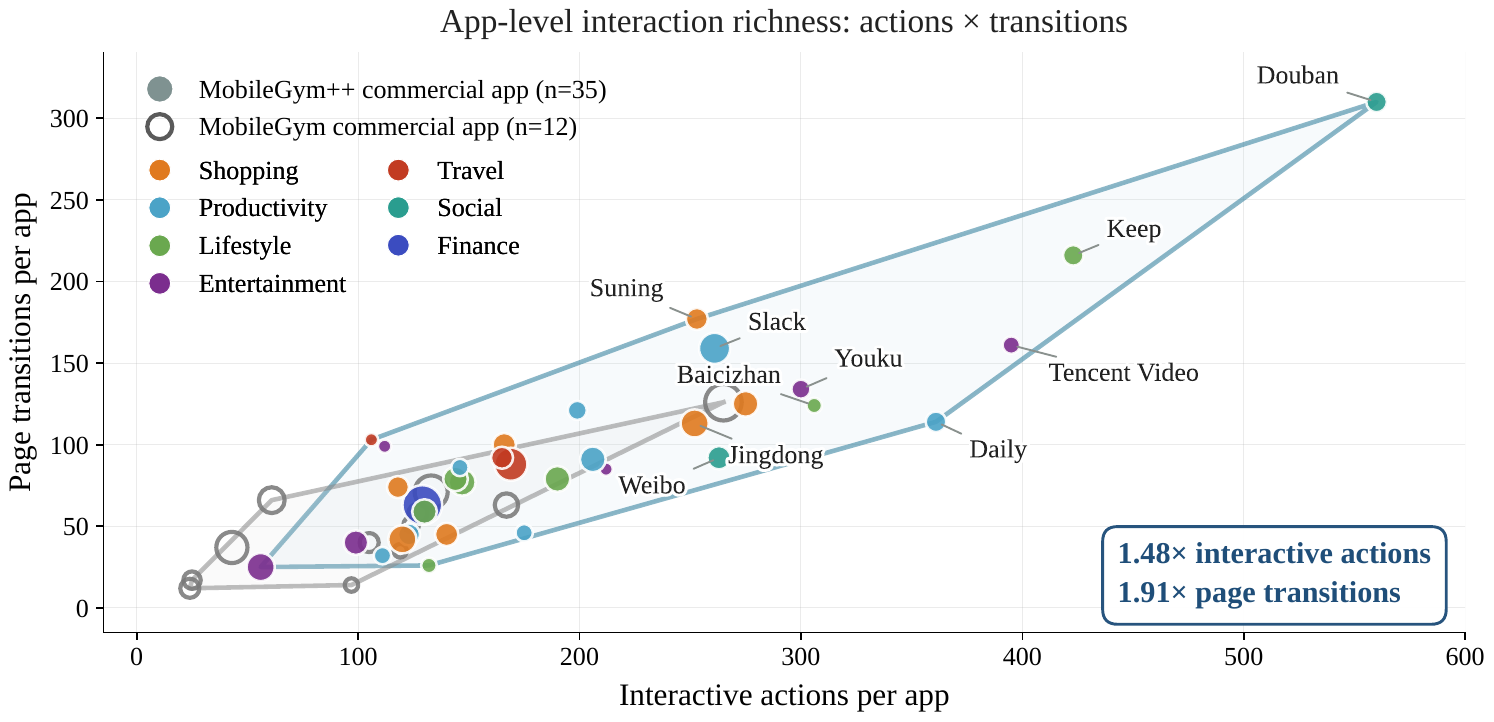}
\caption{
App-level interaction richness. Each marker denotes one app, positioned by its interactive actions and page transitions; marker size reflects the number of Bench tasks using that app. Our new apps reach 1.48$\times$ more actions and 1.91$\times$ more transitions.
}
\label{fig:mgpp-app-scale}
\end{figure}

The 35 applications are structurally faithful replicas of their real-world counterparts. Construction starts from the source application's data schema and interaction graph, specified before any interface is rendered, so business-state richness and navigation depth are architectural properties of each replica. 
Human annotators then iteratively compare each replica against the live application and refine both functionality and visual fidelity until the state transitions, and interface details closely match the reference application (Appendix~\ref{app:mgpp-build}).
Every delivered app is checked in two ways before tasks are written on it: a coverage check maps each declared interaction to its implementation, and a replay of the core flows confirms that each transition produces the declared state change.
Across these clients, the median application exposes 37 routes, 166 interactive actions, and 86 page transitions, with a median interaction depth of 3 from the home screen. Relative to the 12 prior MobileGym business apps retained in the same environment, the newly built clients reach $1.48\times$ the interactive actions and $1.91\times$ the page transitions, as shown in Figure~\ref{fig:mgpp-app-scale}.

\subsection{Realistic cross-application workflows}
\label{sec:mgpp-crossapp}

Many real-world mobile workflows span multiple applications, requiring agents to retrieve information in one app, carry it across application boundaries, and complete an action in another. Single-application benchmarks cannot capture this behavior.

MobileGym++ extends simulated mobile use with explicit cross-application interaction mechanisms, including clipboard transfer, app-to-app sharing, payment redirection, and media selection.
All applications run on a shared, resettable device whose full state can be serialized and restored across episodes. Each app maintains its own business objects---such as orders, expenses, and itineraries---which evolve through interaction. These mechanisms allow information and actions to flow across application boundaries, enabling realistic multi-app workflows rather than isolated single-app tasks.

\subsection{Hybrid GUI and app-native tool use}
\label{sec:mgpp-tools}

Computer use is shifting beyond GUI-only interaction toward hybrid graphical and code interfaces.
We bring this idea to mobile use through an experimental setting in which applications selectively expose app-native tools alongside their GUIs. 
Unlike prior mobile settings with external tools~\citep{kong2026mobileworld}, our tools expose native app functions and operate on the same state as the GUI.

We follow a simple, principled design for tool construction: each tool exposes a structured lookup or write operation over an application's own business objects, while preserving semantic equivalence with the corresponding GUI interaction. A tool call and its GUI counterpart therefore modify the same persistent state and are evaluated by the same terminal condition.
This enables controlled comparison of the two interaction channels without changing tasks, initialization, or success criteria.
We implement over 300 such tools across more than 50 applications, including system clients such as Clock and Notes; tool depth varies by app, from a few search and lookup functions to more than ten read and write operations. All benchmark tasks remain solvable through GUI-only interaction, while hybrid agents can interleave GUI actions and app-native tool calls within the same task.

\subsection{\textsc{MobileGym++ Bench}}
\label{sec:mgpp-bench}

MobileGym++ Bench enables reliable evaluation of mobile agents in commercial-style applications through resettable environments and programmatic verification over application-level business states. Unlike existing benchmarks that rely on device-level signals~\citep{rawles2025androidworld,kong2026mobileworld}, MobileGym++ Bench evaluates whether agents achieve the intended changes in application business objects. Furthermore, the same task suite supports both GUI-only and hybrid GUI–tool interaction, enabling controlled comparison between the two settings. See details in Appendix~\ref{app:mgpp-bench}.

\textbf{Task Design.}
The primary source of difficulty is commercial navigation and information flow. A single-app task requires the agent to locate, filter, and fill fields across a deep page stack before a business object is written. A cross-application task further requires a live value from another application's store to be identified, carried across the application boundary, and written into the correct field. Instructions are phrased as user goals and do not name widgets, tools, or checker fields; the terminal is a write in the application's own business state.

\textbf{Scale and Composition.}
The benchmark comprises 215 tasks across seven business domains, including 109 single-app, 74 two-app, 27 three-app, and 5 four-app tasks. Tasks are balanced across applications and business objects to avoid domination by a small subset of clients.

\textbf{Evaluation Protocol.}
Each task specifies a user goal, a bound initial state, and a programmatic checker over the before/after diff of application business state. The checker does not inspect the trajectory or screenshots and does not rely on a model judge. GUI-only and Hybrid are evaluated on the same tasks, initial states, and checks. The primary metric is success rate.

\section{OpenMobile-Data-v2}
\label{sec:data}

We release \textbf{OpenMobile-Data-v2}: 11.6K demonstration trajectories and more than 2K executable RL tasks collected across our Android emulator environments and \textsc{MobileGym++}. The 11.6K demonstrations are task--trace pairs for imitation, including 1.7K trajectories that interleave GUI actions with app-native tool calls. Each RL task specifies a goal, an initial state, and a programmatic reward, and can be rolled out in MobileGym++ in either GUI-only or hybrid mode.
\subsection{Construction Pipeline}
\label{sec:data-pipeline}
\label{sec:data-collect}

\paragraph{Environment abstraction and task construction.}
We construct a state--action transition graph for each application to expose its available functionality and support the generation of executable user goals following OpenMobile~\citep{cheng2026openmobile}. Nodes represent representative application states, while edges capture actions that induce state transitions. On the Android emulator, we explore each application to discover screens, actions, and transitions, and consolidate perceptually similar screenshots into representative states. This exploration yields a connected view of the application's functionality: the global view captures the overall state--action structure and available functionality, while the local view captures the actions and states reachable from a particular state. For \textsc{MobileGym++}, the same views are derived directly from the backend specification, which explicitly defines pages, actions, and business objects. Conditioned on these views, Gemini-3.1-Pro-Preview generates realistic and executable user instructions. We filter the generated instructions for quality and diversity, removing low-quality and highly similar tasks.

\paragraph{Trajectory collection and filtering.}
Conditioned on the generated user instructions, Gemini-3.1-Pro-Preview acts as the teacher agent and executes each task in its corresponding environment. Each execution produces a trajectory of screens, GUI actions, and, where available, native tool calls. We retain only rollouts that complete the specified task and pass the corresponding runtime-specific verifier: completion is assessed by a vision-language judge on the Android emulator~\citep{sun2026osreward}, while \textsc{MobileGym++} uses its programmatic task verifier. After collecting successful trajectories, we perform step-level filtering to remove redundant interactions, including meaningless action loops and actions that leave the application state unchanged. The resulting data are released as two separate resources: demonstration trajectories retain the full observation--action sequence, while executable RL tasks retain only the instruction, initial state, and verifier.

\subsection{Data Statistics}
\label{sec:data-quality}
\label{sec:data-stats}
\begin{figure}[t]
\centering
    \includegraphics[
        width=\linewidth
    ]{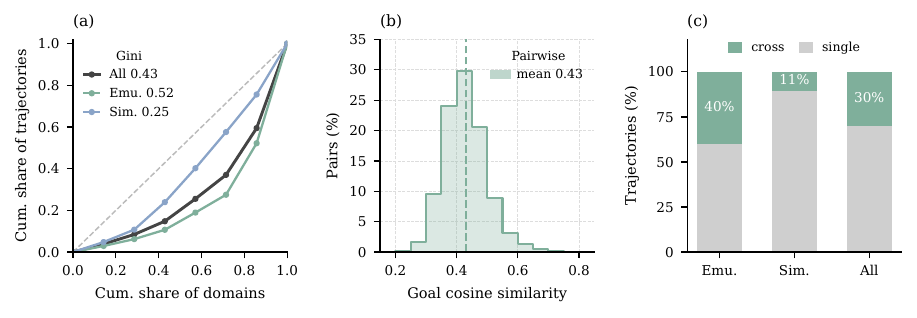}
\caption{Diversity and coverage of OpenMobile-Data-v2.
\textbf{(a)}~Domain distribution across seven categories for emulator and simulator data. Gini coefficients quantify domain balance.
\textbf{(b)}~Pairwise cosine similarity among task goals.
\textbf{(c)}~Single-app and cross-app workflow composition.}
\label{fig:data-quality}
\end{figure}

OpenMobile-Data-v2 is diverse along three dimensions: domain distribution,
task diversity, and workflow diversity (Figure~\ref{fig:data-quality}).
The data span seven domains with a relatively balanced distribution. We quantify domain balance using the Gini coefficient, where lower values indicate more uniform coverage. The full set achieves a Gini coefficient of 0.43, while the emulator subset is more concentrated (0.52) and simulated clients provide more balanced coverage (0.25).
Task goals also show diversity, with a mean pairwise cosine
similarity of 0.43, indicating varied requests without excessive repetition.
Finally, the data cover both single-app and cross-app workflows, with 30\%
of trajectories spanning multiple applications (40\% on the emulator and
11\% on simulated clients). Together, these statistics show that
OpenMobile-Data-v2 provides diverse domains, task goals, and interaction
workflows.

\section{Experiments}
\label{sec:exp}

In this section, we evaluate models trained with OpenMobile-Data-v2 on established mobile-agent benchmarks, real-device tasks, and \textsc{MobileGym++}. 

\subsection{Training and Evaluation Details}
\label{sec:exp-details}

We train two supervised fine-tuning runs from each base model, both for 5 epochs. GUI-only SFT uses only GUI trajectories. Hybrid SFT uses the full mixture, combining those GUI trajectories with hybrid GUI--tool trajectories. RL is then initialized from the hybrid SFT checkpoint and trained on 512 tasks randomly sampled from the 2K+ executable tasks. We evaluate the resulting models on five benchmarks spanning three settings---emulator apps, physical devices, and simulated apps with both GUI-only and tool-augmented interaction. \textbf{AndroidWorld}~\citep{rawles2025androidworld} contains 116 tasks across 20 real apps on the emulator, with success determined by the resulting device state; \textbf{MobileWorld}~\citep{kong2026mobileworld} contains 201 harder long-horizon and cross-app tasks evaluated in GUI-only and MCP-enabled settings; \textbf{MobileGym}~\citep{wu2026mobilegym} uses resettable mock apps with state-based verification and reports success rate (SR) and progress rate (PR); \textbf{\textsc{MobileGym++} Bench} (Section~\ref{sec:mgpp-bench}) contains 215 single- and cross-app tasks evaluated with identical tasks and checkers in GUI-only and hybrid settings; and \textbf{SpaBench}~\citep{chen2024spa} evaluates tasks on physical Android devices, providing a real-device evaluation beyond the emulator-based training environments. Detailed training and evaluation configurations are provided in (Appendix~\ref{app:details}).

We analyze the overlap between our training data and benchmark tests to
assess potential benchmark overfitting (Appendix~\ref{app:leakage}).
Using \texttt{text-embedding-3-large}, only 1.11\%, 0.39\%, 0.46\%, and
2.25\% of SFT instructions exceed a cosine similarity of 0.7 with
AndroidWorld, MobileWorld, MobileGym, and \textsc{MobileGym++} Bench,
respectively, and none exceeds 0.9. Among the 2,475 RL task goals, 6.1\%
exceed 0.7 with \textsc{MobileGym++} Bench, while none exceeds 0.9.

\subsection{Main results}
\label{sec:exp-main}

\begin{table*}[h]
\centering
\caption{\textbf{Main results} on established mobile agent benchmarks and real-device evaluation.
$*$ denotes results reproduced by us, and
$\pm$ denotes the standard deviation across evaluation seeds.}
\label{tab:main}
\small
\resizebox{\textwidth}{!}{
\begin{tabular}{lccccccccc}
\toprule
Method &
\multicolumn{2}{c}{AndroidWorld} &
\multicolumn{2}{c}{MobileWorld} &
\multicolumn{2}{c}{MobileGym} &
\multicolumn{2}{c}{MobileGym++} &
SPA-Bench \\
\cmidrule(lr){2-3}
\cmidrule(lr){4-5}
\cmidrule(lr){6-7}
\cmidrule(lr){8-9}
&
Pass@1$\uparrow$ & Pass@3$\uparrow$
& GUI & MCP
& SR & PR
& GUI & Hybrid
& L3 \\
\midrule
\multicolumn{10}{c}{\textit{Commercial \& Closed-Weight Models}} \\
\midrule
GPT-5.6-Sol
  & -- & --
  & 70.1 & --
  & -- & --
  & -- & --
  & 68.1* \\
Gemini-3.1-pro-preview
  &70.7 & --
  & 58.1 & --
  & -- & --
  & -- & --
  & 70.2* \\
MAI-UI-32B
  & 73.3 & --
  & 36.2 & 30.0
  & -- & --
  & -- & --
  & -- \\
MAI-UI-235B-A22B
  & 76.7 & --
  & 39.7 & 37.5
  & -- & --
  & -- & --
  & -- \\
\midrule
\multicolumn{10}{c}{\textit{Open-Weight Models}} \\
\midrule
UI-Venus-1.5-8B
  & 73.7 & --
  & -- & --
  & 15.4 & 28.3
  & 19.5* & 16.3*
  & 29.8* \\
GUI-Owl-1.5-8B
  & 69.0 & --
  & 38.2 & 37.5
  & 17.2* & 31.3*
  & 13.0* & 12.6*
  & -- \\
GUI-Owl-1.5-32B
  & 69.4 & --
  & 43.9 & 42.5
  & 18.4* & 30.9*
  & 18.6* & 23.3*
  & -- \\
MAI-UI-8B
  & 70.7 & --
  & 27.5 & 20.0
  & 16.8* & 28.3*
  & 19.1* & 18.6*
  & -- \\
\midrule
\multicolumn{10}{c}{\textit{Open-Data Models}} \\
\midrule
OpenMobile-1-8B
  & 64.7 $\pm$ 3.2 & 78.0
  & 17.7 & --
  & 13.7* & 25.5*
  & 13.5* & 7.0*
  & -- \\
\multicolumn{10}{l}{OpenMobile-2-9B} \\
Base (Qwen3.5-9B)
  & 54.3 $\pm$ 3.7 *  & 69.8*
  & 12.8* & 12.5*
  & 9.4* & 20.0*
  & 9.3* & 12.1*
  & 25.5* \\
\quad GUI-only SFT
  & 72.8 $\pm$ 2.2 & 81.9
  & 38.5 & 12.5
  & 34.8 & 51.6
  & 40.9 & 41.9
  & -- \\
\quad Hybrid SFT
  & 73.3 $\pm$ 1.5 & 83.6
  & 39.3 & 27.5
  & 35.9 & 52.2
  & 39.1 & 44.7
  & 48.9 \\
\quad+ RL
  & 74.0 $\pm$ 2.9 & 83.6
  & 38.5 & 27.5
  & 38.3 & 52.0
  & 40.5 & 46.5
  & 48.9 \\
\addlinespace
\multicolumn{10}{l}{OpenMobile-2-27B} \\
Base (Qwen3.6-27B)
  & 67.1 $\pm$ 4.1 *& 81.9*
  & 35.9* & 32.5*
  & 37.1* & 50.1*
  & 46.5* & 50.7*
  & 31.9* \\
\quad GUI-only SFT
  & 80.8 $\pm$ 2.1 & 89.7
  & 50.4 & 35.0
  & 46.9 & 63.1
  & 54.9 & 57.7
  & -- \\
\quad Hybrid SFT
  & 80.8 $\pm$ 3.5 & 90.5
  & 48.7 & 45.0
  & 48.4 & 64.5
  & 55.3 & 65.6
  & 57.4 \\
\quad + RL
  & 79.9 $\pm$ 3.0 & 90.5
  & 50.4 & 42.5
  & 51.2 & 66.6
  & 57.2 & 67.0
  & 59.6 \\
\bottomrule
\end{tabular}
}
\end{table*}

\paragraph{OpenMobile-Data-v2 improves performance across established mobile agent benchmarks.}
As shown in Table~\ref{tab:main}, models fine-tuned on OpenMobile-Data-v2 consistently outperform their corresponding base models across AndroidWorld, MobileWorld, and MobileGym++. The improvements are substantial at both model scales: the 9B model gains 19.7 points on AndroidWorld over its base (54.3 to 74.0), while the 27B model gains 12.8 points (67.1 to 79.9), with consistent improvements on MobileWorld and MobileGym++ as well. These results show that the interaction experience synthesized by OpenMobile-Data-v2 is useful across different mobile environments.

Starting from hybrid SFT, we further explore agentic RL on executable mobile tasks. 
RL improves training-aligned environments, raising MobileGym success rate from 35.9 to 38.3 (9B) and from 48.4 to 51.2 (27B), while also improving 27B SPA-Bench performance from 57.4 to 59.6, suggesting improved transfer to real-device mobile use. Gains on other benchmarks remain limited, suggesting that RL benefits depend on the alignment between training environments and evaluation scenarios.

\vspace{-2mm}
\paragraph{Generalization beyond simulated environments to real devices.}
Beyond established online benchmarks, we evaluate the resulting agents on real Android devices using SPA-Bench L3. Hybrid SFT achieves 48.9 and 57.4 for the 9B and 27B models, respectively, substantially improving over their corresponding base models (25.5 and 31.9) and outperforming strong open-source counterparts. These results show that the benefits of OpenMobile-Data-v2 transfer beyond the controlled environments used for training and evaluation to physical devices.

\vspace{-2mm}
\paragraph{App-native tools provide a complementary action channel.}
MobileGym++ enables agents to interleave GUI actions with application-specific tool calls within a task. On MobileGym++ Bench, hybrid SFT improves the hybrid evaluation over GUI-only SFT at both scales (9B: 41.9 to 44.7; 27B: 57.7 to 65.6), while GUI-only success stays comparable (Table~\ref{tab:main}). This result demonstrates that app-native tools provide a useful complementary action channel for mobile agent interaction.

\vspace{-2mm}
\paragraph{Hybrid training enables generalizable tool-use behavior.}
We compare hybrid and GUI-only SFT to evaluate whether app-native tool training transfers beyond the observed tools. On MobileWorld MCP, whose tools are unseen during training, hybrid SFT achieves 27.5 and 45.0 versus 12.5 and 35.0 for GUI-only SFT. This transfer suggests that hybrid training learns general tool-use strategies rather than memorizing specific tool APIs.

\label{sec:exp-gui}

\subsection{Analysis}
\label{sec:analysis}
\begin{figure*}[h]
\centering
\begin{minipage}[t]{0.40\textwidth}
\centering
\includegraphics[
    width=\linewidth,
    keepaspectratio
]{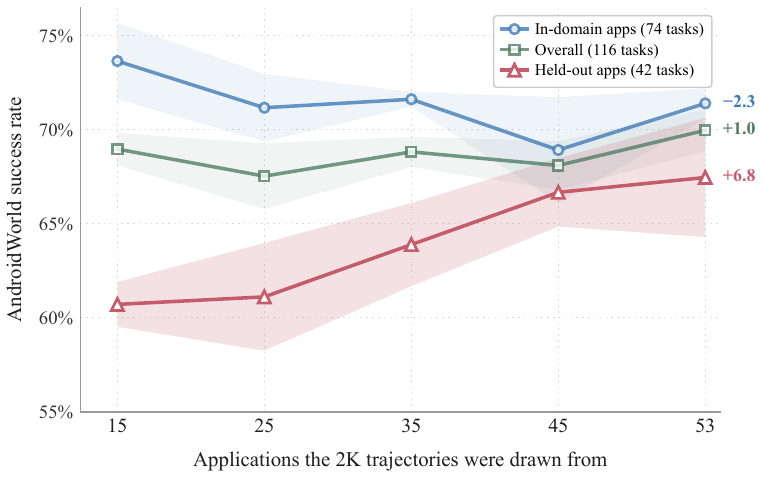}\\[-0.2em]
{\small (a) Environment coverage}
\end{minipage}
\hfill
\begin{minipage}[t]{0.56\textwidth}
\centering
\includegraphics[
    width=\linewidth,
    keepaspectratio
]{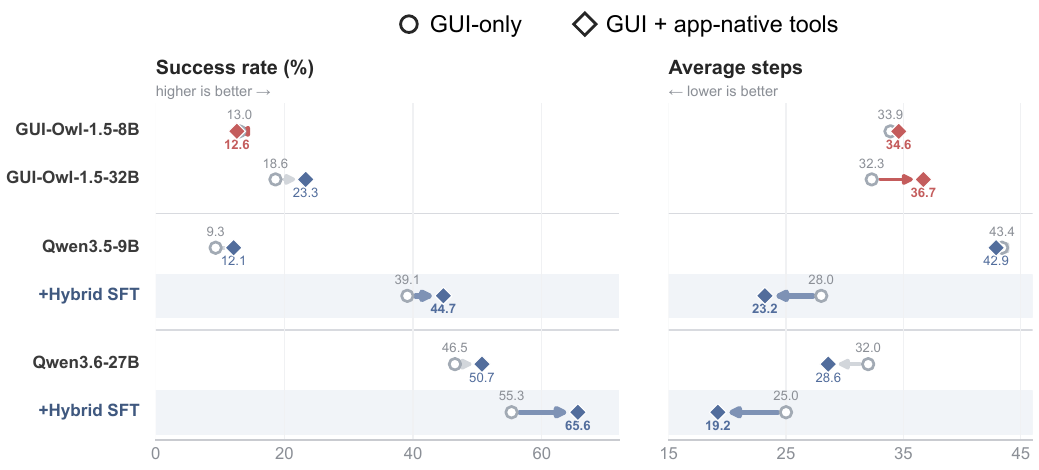}\\[-0.2em]
{\small (b) App-native tool use}
\end{minipage}
\caption{
\textbf{Environment coverage and app-native tool use.}
(a)~Broader coverage improves held-out AndroidWorld performance at a fixed 2K-trajectory budget.
(b)~App-native tools improve success and efficiency, with larger gains after Hybrid SFT.
}
\label{fig:analysis-results}
\end{figure*}

We focus our analysis on two aspects of mobile agent interaction: how environment coverage relates to generalization, and how app-native tools complement GUI interaction.

\paragraph{Broader environment coverage improves transfer under a fixed data budget.}
With the backbone, optimizer, and total training data fixed at 2K emulator
trajectories, we vary application coverage from 15 to 53 while keeping the
same 15 InD applications. Broader coverage improves transfer to unseen
applications, with Pass@1 increasing from 60.7 to 67.5 (Figure~\ref{fig:analysis-results}a).
Meanwhile, the InD data share drops from 100\% to 39.1\%, but InD Pass@1
remains largely stable (73.7$\rightarrow$71.4; Table~\ref{tab:env-split}).
These results show that broader application coverage improves generalization
without significantly compromising performance on familiar applications.

\begin{table}[h]
\centering
\small
\begin{minipage}[t]{0.33\linewidth}
\centering
\caption{
Environment-coverage settings under a fixed budget of 2K trajectories. InD share denotes the fraction of training trajectories from the 15 in-distribution applications.
}
\label{tab:env-split}
\setlength{\tabcolsep}{3pt}
\begin{tabular}{cc}
\toprule
\# Train apps & InD data share \\
\midrule
15 & 100\% \\
25 & 64.0\% \\
35 & 48.9\% \\
45 & 41.1\% \\
53 & 39.1\% \\
\bottomrule
\end{tabular}
\end{minipage}
\hfill
\begin{minipage}[t]{0.63\linewidth}
\centering
\caption{App-native tool use on MobileGym++.
TR (tool-call rate) is the average number of app-native tool calls per task. \#Tools is the number of distinct tools invoked across the benchmark, and \#Tasks is the number of tasks (out of 215) in which at least one app-native tool is called.}

\label{tab:tool_use}
\setlength{\tabcolsep}{3pt}
\resizebox{0.76\linewidth}{!}{%
\begin{tabular}{llccc}
\toprule
Model & Stage & TR & \#Tools & \#Tasks \\
\midrule
GUI-Owl-1.5-8B &- &0.58&15&20\\
GUI-Owl-1.5-32B &- &0.51&51&45\\
\midrule
\multirow{2}{*}{Qwen3.5-9B} & Base & 0.64 & 28 & 21 \\
 & + Hybrid SFT & 1.41 & 110 & 113 \\
\midrule
\multirow{2}{*}{Qwen3.6-27B} & Base & 1.64 & 97 & 102 \\
 & + Hybrid SFT & 2.23 & 115 & 128 \\
\bottomrule
\end{tabular}%
}
\end{minipage}
\end{table}

\paragraph{Hybrid training unlocks the benefits of app-native tools.}
Providing app-native tools alone does not guarantee effective tool use: without hybrid training, models rarely learn when to invoke them.
Before hybrid SFT, tool calls appear in only 20--102 of 215 tasks (Table~\ref{tab:tool_use}), bringing limited gains: Qwen models improve by only 2.8--4.2 points, while GUI-Owl-1.5 shows inconsistent results and can even degrade (Figure~\ref{fig:analysis-results}b).
After hybrid SFT, tool usage becomes substantially more effective: tool calls appear in 113 and 128 tasks, covering 110 and 115 distinct tools for the 9B and 27B models, respectively. More importantly, tool-enabled gains increase to 5.6 and 10.3 points, while trajectories become 4.8 and 5.8 steps shorter. These results show that hybrid training teaches models to effectively leverage app-native tools, improving both task success and interaction efficiency.

\section{Conclusion}
\label{sec:conclusion}

We presented OpenMobile-2, an open training stack for mobile agents spanning
diverse environments, open training data, and trained agent models. On the environment side, \textsc{MobileGym++} and newly configured emulator apps establish a diverse playground spanning over 110 applications, including high-fidelity commercial-style apps with clean reset and verification.
On the data side, we release nearly 12K demonstration trajectories and over 2K executable
RL tasks with automatic rewards. Models trained on this stack demonstrate
strong performance across established mobile agent benchmarks and transfer to real-device evaluation. 
We further explore app-native tools as a complementary action interface for mobile agents, finding that tool access alone brings limited gains while hybrid training substantially amplifies their benefits.
We open-source environments, data, benchmarks, and trained models to support further open research on mobile agents.


\bibliography{iclr2027_conference}
\bibliographystyle{iclr2027_conference}

\appendix
\newpage
\section{Training and Evaluation Details}
\label{app:details}

\subsection{Supervised fine-tuning}
\label{app:sft}

\paragraph{Two SFT runs.}
Supervised fine-tuning consists of two separate 5-epoch runs from the base model. They share the hyperparameters in Table~\ref{tab:app-hparams} and differ in training data (Table~\ref{tab:app-sft-data}). The GUI-only run uses emulator trajectories together with GUI simulator trajectories, and does not use app-native tools. The hybrid run uses the full emulator and simulator mixture, including 1,774 simulator trajectories that interleave app-native tool calls. GUI trajectories use the GUI-only system prompt, whereas hybrid trajectories use a tool-aware system prompt. RL is initialized from the hybrid-run checkpoint.

\paragraph{Training samples.}
Each trajectory yields one training sample per step. A sample contains the task instruction, the interaction history up to that step, and a target turn consisting of a \texttt{Thought:} paragraph followed by one tool call. The three most recent screenshots are retained as images; older screenshots are replaced with the text ``This screenshot has been collapsed.'' The full mixture contains 118,582 step-level samples. SFT screenshots are fed at their native $1080\times2400$ resolution, and the model predicts coordinates in a normalized $999\times999$ space.

\paragraph{Action space.}
The GUI-only run uses the \texttt{mobile\_use} tool with actions \texttt{click}, \texttt{long\_press}, \texttt{swipe}, \texttt{type}, \texttt{answer}, \texttt{system\_button} (Back, Home, Menu, or Enter), \texttt{wait}, and \texttt{terminate} (success or failure). The hybrid run adds app-native tools, including actions such as \texttt{clock\_\_create\_alarm} and \texttt{notes\_\_create\_note}. An application's tool catalog is available only while that application is in the foreground, so the policy chooses between a GUI action and an available app-native action at each step.

\begin{table}[H]
\centering
\small
\caption{Training hyperparameters. SFT values are shared by both runs; -- indicates that a setting does not apply.}
\label{tab:app-hparams}
\setlength{\tabcolsep}{3pt}
\begin{tabular}{p{0.27\linewidth}p{0.32\linewidth}p{0.32\linewidth}}
\toprule
Hyperparameter & SFT & RL \\
\midrule
Framework & Megatron-SWIFT & verl \\
Objective & Next-token prediction & DAPO (group-relative) \\
Trainable parameters & LLM (vision frozen) & LoRA \\
Optimizer & Adam ($\beta_1{=}0.9$, $\beta_2{=}0.95$) & AdamW \\
Weight decay / gradient clip & 0.1 / 1.0 & 0.1 / 1.0 \\
Learning rate & $5\times10^{-6}$ & $1\times10^{-6}$ \\
Schedule & Cosine; 3\% warmup; minimum $10^{-7}$ & 40 warmup steps \\
Batch size & 32 (microbatch 1) & 64 tasks $\times$ 8 rollouts \\
Training length & 5 epochs & 150 steps (budget 300); 2 updates each \\
Group size & -- & 8 \\
Clipping $\epsilon_{\text{low}}$ / $\epsilon_{\text{high}}$ & -- & 0.2 / 0.28 \\
KL coefficient & -- & 0.01 (low-variance estimator) \\
Token budget & 32,768 (packed) & 32,768 (4,096 prompt + 28,672 response; 1,024 per turn) \\
Screenshot input & $1080\times2400$ (native) & Short side 448 px \\
Screenshot history & Last 3 & Full episode (at most 32 images) \\
Coordinate space & $999\times999$ & $999\times999$ \\
Step cap & None (maximum observed: 50) & 30 \\
Precision & bf16 & bf16 \\
Hardware & 8 GPUs; tensor parallelism 4 & 32 H800 GPUs; sequence parallelism 4 \\
\bottomrule
\end{tabular}
\end{table}

\begin{table}[t]
\centering
\small
\caption{Trajectories in each SFT run. The hybrid run adds 1,774 simulator trajectories that interleave app-native tool calls, so the simulator count rises from 2,144 to 3,918.}
\label{tab:app-sft-data}
\setlength{\tabcolsep}{5pt}
\begin{tabular}{lrrrr}
\toprule
 & \multicolumn{2}{c}{GUI-only run} & \multicolumn{2}{c}{Hybrid run} \\
\cmidrule(lr){2-3} \cmidrule(lr){4-5}
Source & Trajectories & Steps & Trajectories & Steps \\
\midrule
Emulator & 7,725 & 83,631 & 7,725 & 83,631 \\
Simulator & 2,144 & 22,857 & 3,918 & 34,951 \\
\midrule
Total & 9,869 & 106,488 & 11,643 & 118,582 \\
\bottomrule
\end{tabular}
\end{table}

\subsection{Reinforcement learning}
\label{app:rl}

\paragraph{Algorithm.}
We initialize reinforcement learning from the hybrid SFT checkpoint and optimize the policy with GRPO-style group-relative advantages~\citep{shao2024deepseekmath} and the DAPO objective~\citep{yu2026dapo}, implemented in verl~\citep{sheng2025hybridflow}. For each task, we sample a group of 8 rollouts in \textsc{MobileGym++} and use asymmetric clipping with $\epsilon_{\text{low}}{=}0.2$ and $\epsilon_{\text{high}}{=}0.28$. We add a KL loss toward the SFT reference policy, with coefficient 0.01 and a low-variance estimator. The hyperparameters are listed in Table~\ref{tab:app-hparams}.

\paragraph{Tasks.}
Training uses all 512 tasks. Each task has a bound initial state and a programmatic verifier in \textsc{MobileGym++}. Validation uses the official MobileGym split of 256 tasks, run once before training and every 10 steps, with one rollout per task.

\paragraph{Rollouts.}
Each rollout is a multi-turn episode in the hybrid setting. Every assistant turn contains a \texttt{Thought:} paragraph and one tool call, with at most 1,024 generated tokens per turn. Unlike SFT, the screenshot history of the episode remains in context, resized so that the short side is 448 pixels and capped at 32 images. The action space contains nine \texttt{mobile\_use} actions (including \texttt{open\_app}) and app-native tools. An application's tool catalog is provided when it comes to the foreground; each tool call returns a structured receipt containing its output or an error code. An episode ends when the agent calls \texttt{terminate} or reaches the smaller of the task-specific and global step limits (Table~\ref{tab:app-hparams}).

\paragraph{Reward.}
The reward of a rollout is
\begin{equation*}
r = \underbrace{v}_{\text{verifier}} \cdot \underbrace{m_{\text{tool}}}_{\text{tool bonus}} \cdot \underbrace{\ell}_{\text{length factor}} - 0.2\, n_{\text{fmt}} - 0.2\, f_{\text{succ}}.
\end{equation*}
Here $v\in\{0,1\}$ is a relaxed state verifier: the goal is met, and any extra state changes are benign. Reported accuracy uses the strict verifier, which also requires a clean state. In the hybrid setting, $m_{\text{tool}}=2$ when the trajectory contains at least one app-native tool call, and $m_{\text{tool}}=1$ otherwise. The length factor $\ell$ is 1 for responses up to 12,000 tokens and decreases linearly to 0.5 at 28,672 tokens. $n_{\text{fmt}}$ is the number of malformed turns, and $f_{\text{succ}}=1$ when the agent claims completion but the verifier fails.

\paragraph{Infrastructure.}
Training runs on 32 H800 GPUs across 4 nodes. The actor is a rank-64 LoRA ($\alpha{=}64$) on the attention and MLP projections, with the vision tower frozen, trained in bf16 with FSDP2 and Ulysses sequence parallelism of degree 4. No parameters, optimizer states, or activations are offloaded. Rollouts are generated by a colocated vLLM engine with tensor parallelism 4. Each node hosts a \textsc{MobileGym++} simulator, and rollouts drive 64 browser pages concurrently (32 workers, two pages each). The learning rate is $1\times10^{-6}$ after 40 warmup steps, with an entropy coefficient of 0.001.

\subsection{Evaluation benchmarks}
\label{app:eval-bench}

We evaluate on five benchmarks that differ in where the apps run, which actions the agent may take, and how completion is judged, so a gain that holds on all of them is unlikely to come from fitting one environment.

\textbf{AndroidWorld}~\citep{rawles2025androidworld} is a widely adopted benchmark for evaluating mobile agents in reproducible Android environments. It comprises 116 programmatic tasks across 20 real-world apps, with deterministic evaluation based on the resulting device state. We report task success rate.

\textbf{MobileWorld}~\citep{kong2026mobileworld} is a recent benchmark designed to evaluate more challenging mobile-agent capabilities. It contains 201 tasks across 20 apps, with an emphasis on long-horizon and cross-application workflows. We evaluate both its GUI-only tasks and MCP-enabled tasks using the official task checker.

\textbf{MobileGym}~\citep{wu2026mobilegym} is a benchmark built around resettable mock applications with programmatic state-based evaluation. It evaluates agents on tasks requiring application interactions and state transitions, and reports both success rate (SR) and progress rate (PR), capturing complete and partial task completion, respectively.

\textbf{\textsc{MobileGym++} Bench} (Section~\ref{sec:mgpp-bench}) is designed to evaluate mobile agents on tasks involving application business state and native tool interactions. It contains 215 tasks spanning single-app and cross-app workflows, each initialized from a fixed application state and evaluated by a programmatic checker over the application business state. We evaluate both GUI-only and hybrid GUI and app-native tool interaction on the same tasks, initial states, and checker.

\textbf{SpaBench}~\citep{chen2024spa} evaluates mobile agents on real Android devices, providing an evaluation setting that goes beyond the controlled emulator and simulator environments used in training and the other benchmarks. We use SpaBench to assess whether the capabilities learned from OpenMobile-2 transfer to real-device interaction.

\subsection{Overlap with evaluation benchmarks}
\label{app:leakage}

\begin{figure}[h]
\centering
\begin{minipage}[t]{0.49\linewidth}
\centering
\includegraphics[width=\linewidth]{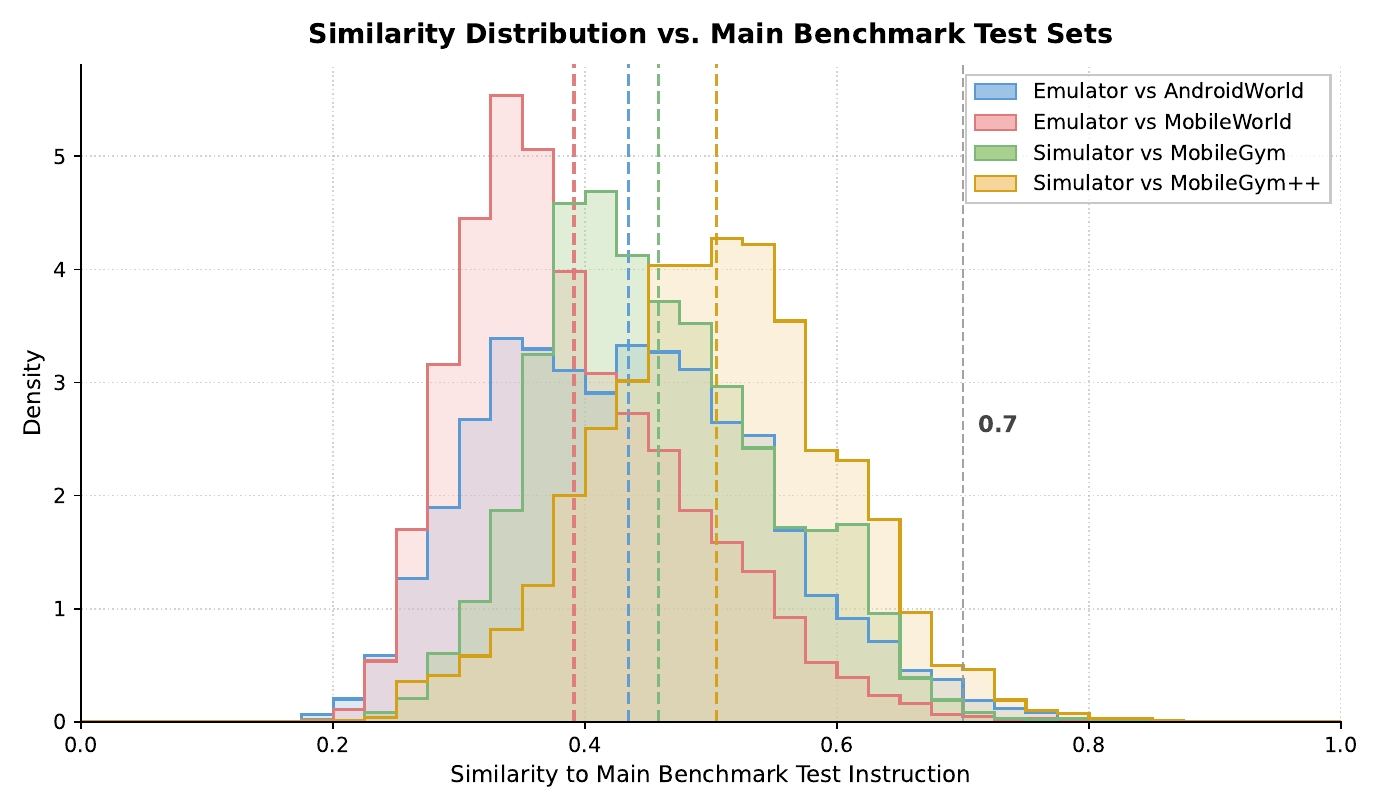}\\[-0.2em]
{\small (a) SFT goals}
\end{minipage}\hfill
\begin{minipage}[t]{0.49\linewidth}
\centering
\includegraphics[width=\linewidth]{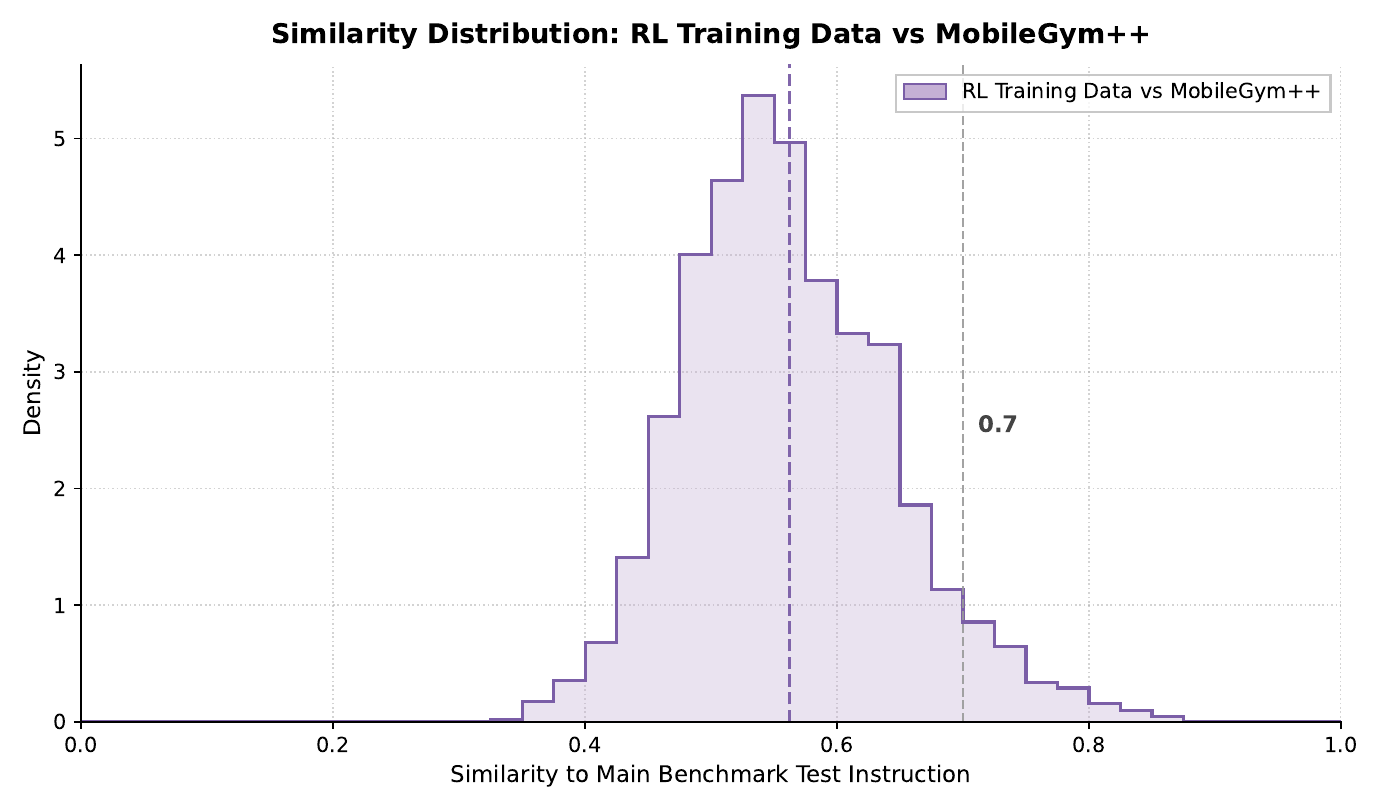}\\[-0.2em]
{\small (b) RL goals}
\end{minipage}
\caption{Per-goal maximum cosine similarity. (a)~SFT goals against the matched benchmark. (b)~RL task goals against \textsc{MobileGym++} Bench. Vertical lines mark each mean and the 0.7 threshold.}
\label{fig:leakage}
\end{figure}
We compare training goals with the evaluation instructions of the four benchmarks that share a runtime with training. Emulator goals (7,718 unique) are compared with AndroidWorld and MobileWorld; simulator goals (3,916 unique) are compared with MobileGym and \textsc{MobileGym++} Bench. Each goal is embedded with \texttt{text-embedding-3-large} (3,072 dimensions, L2-normalized), and we record its maximum cosine similarity to that benchmark. Table~\ref{tab:leakage} and Figure~\ref{fig:leakage}a summarize these scores. No training goal is an exact string match of an evaluation instruction, and none exceeds similarity 0.9. 

\begin{table}[h]
\centering
\small
\caption{Maximum cosine similarity of each SFT training goal to the matched benchmark. Above 0.7 counts goals whose maximum exceeds 0.7.}
\label{tab:leakage}
\begin{tabular}{lrrrr}
\toprule
Benchmark & Train goals & Mean & Max & Above 0.7 \\
\midrule
AndroidWorld & 7,718 & 0.434 & 0.892 & 1.11\% (86) \\
MobileWorld & 7,718 & 0.391 & 0.851 & 0.39\% (30) \\
MobileGym & 3,916 & 0.458 & 0.826 & 0.46\% (18) \\
\textsc{MobileGym++} Bench & 3,916 & 0.504 & 0.861 & 2.25\% (88) \\
\bottomrule
\end{tabular}
\end{table}

The same embedding is applied to the 2,475 RL task goals, compared only with \textsc{MobileGym++} Bench (Figure~\ref{fig:leakage}b). Exact string overlap is again zero, and none exceeds 0.9. The mean maximum similarity is 0.562, the maximum is 0.862, and 6.1\% (151) exceed 0.7.

\section{MobileGym++ App Construction}
\label{app:mgpp-build}

\paragraph{Behavior specification.}
We represent core application workflows as a structured graph of page nodes and meaningful UI states. Each state-changing interaction names its UI element and action, describes the expected state effect, and includes a check for the resulting state. This makes navigation and business-state writes explicit before implementation. The specification is then used as the contract for implementing and checking routes, actions, and state changes.

\par\medskip
\noindent
\begin{minipage}{\linewidth}
\centering
\includegraphics[width=\linewidth]{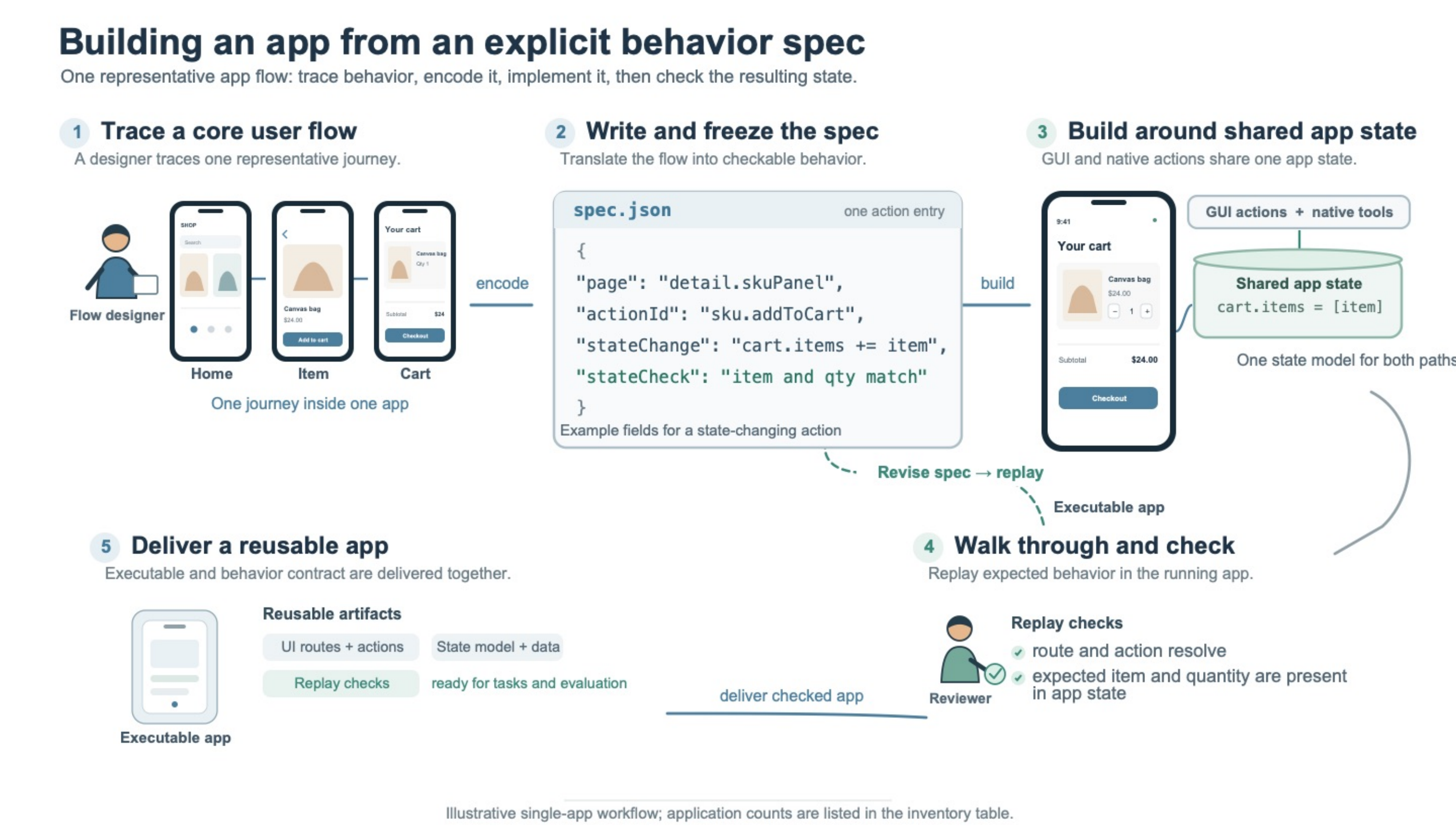}
\captionof{figure}{Illustrative single-app workflow. The phone screens show one journey; the numbered circles mark construction stages, not application counts.}
\label{fig:app-construction-process}
\end{minipage}
\par\medskip

\paragraph{Specification entry.}
The excerpt below shows the form of a state-changing interaction. A page identifier groups its controls; the action identifier names the operation; and the state change and check make the outcome observable.
\begin{minipage}{\linewidth}
\begin{lstlisting}[basicstyle=\ttfamily\small, frame=single, columns=fullflexible, breaklines=true]
{
  "id": "detail.skuPanel",
  "interactions": [{
    "element": "Add to cart",
    "effect": "state",
    "actionId": "sku.addToCart.confirm",
    "stateChange": "cart.items += item",
    "stateCheck": "item and quantity match"
  }]
}
\end{lstlisting}
\end{minipage}

\paragraph{Implementation and checks.}
Routes and actions operate on app-owned state and seeded data. GUI actions and app-native tools use the same app state. A coverage check maps declared interaction identifiers to implementation bindings; runtime replay checks representative transitions and their expected state effects. When a mismatch appears, the specification is corrected and the affected checks are replayed. Each delivered app includes its executable interface, state model, seeded data, and reusable checks for core flows. Benchmark instructions and terminal verifiers are authored separately on top of the completed apps.

\section{Playground Inventory}
\label{app:mgpp-apps}

\textsc{MobileGym++} contains 63 installable simulated clients: 35 applications built in this work and 28 inherited from MobileGym. Of these, 47 are commercial-style applications and 16 are system clients. The Android emulator inventory contains 62 applications. The tables below list applications by domain; the emulator inventory is not divided by source batch.

\begin{table}[H]
\centering
\small
\caption{Application counts by domain in the two runtimes.}
\label{tab:app-domain-counts}
\begin{tabular}{lrr}
\toprule
Domain & \textsc{MobileGym++} & Emulator \\
\midrule
Lifestyle & 18 & 19 \\
Productivity & 15 & 23 \\
Entertainment & 9 & 15 \\
Shopping & 8 & 1 \\
Travel & 5 & 3 \\
Social & 6 & 0 \\
Finance & 2 & 1 \\
\midrule
Total & 63 & 62 \\
\bottomrule
\end{tabular}
\end{table}

\subsection{MobileGym++ applications}
\label{app:mgpp-app-table}

Table~\ref{tab:app-mgpp-list} gives the complete 63-client inventory; $^\ast$ marks the 35 applications built in this work.

\begin{table}[H]
\centering
\small
\caption{The 63 \textsc{MobileGym++} clients. $^\ast$ marks the 35 applications built in this work and $^\dagger$ marks the 16 system clients.}
\label{tab:app-mgpp-list}
\begin{tabular}{lp{0.74\linewidth}}
\toprule
Domain & Applications \\
\midrule
Lifestyle (18) & Baicizhan$^\ast$, Cainiao$^\ast$, Dianping$^\ast$, Keep$^\ast$, Luckin Coffee$^\ast$, Maoyan$^\ast$, Meituan$^\ast$, Meiyou$^\ast$, Weather$^\dagger$, Answer Sheet$^\dagger$, Browser$^\dagger$, Calculator$^\dagger$, Calculator~2$^\dagger$, Clock$^\dagger$, Compass$^\dagger$, Gallery$^\dagger$, Settings$^\dagger$, Themes$^\dagger$ \\
Productivity (15) & BOSS Zhipin$^\ast$, Days Matter$^\ast$, TickTick$^\ast$, Fanqie ToDo$^\ast$, flomo$^\ast$, Google Drive$^\ast$, Mail$^\ast$, Slack$^\ast$, Daily$^\ast$, Tencent Meeting, Calendar$^\dagger$, Phone$^\dagger$, Files$^\dagger$, Notes$^\dagger$, Messages$^\dagger$ \\
Entertainment (9) & Podcasts$^\ast$, Fanqie Novel$^\ast$, Tencent Video$^\ast$, NetEase Cloud Music$^\dagger$, Ximalaya$^\ast$, Youku$^\ast$, Bilibili, Spotify, WeRead \\
Shopping (8) & JD$^\ast$, Pinduoduo$^\ast$, Suning$^\ast$, Taobao$^\ast$, Taobao Instant Commerce$^\ast$, Vipshop$^\ast$, Xianyu$^\ast$, eBay \\
Travel (5) & Ctrip$^\ast$, DiDi$^\ast$, Qunar$^\ast$, Maps, Railway 12306 \\
Social (6) & Douban$^\ast$, Weibo$^\ast$, RedNote, Reddit, WeChat, X \\
Finance (2) & Alipay, Kapi$^\ast$ \\
\bottomrule
\end{tabular}
\end{table}

\subsection{Emulator applications}
\label{app:emu-apps}

\begin{table}[H]
\centering
\small
\caption{The 62 Android emulator applications, grouped by domain. AndroidWorld and MiniWoB denote the task-support apps installed with the AndroidWorld suite.}
\label{tab:app-emu-list}
\begin{tabular}{lp{0.74\linewidth}}
\toprule
Domain & Applications \\
\midrule
Lifestyle (19) & Camera, Chrome, Clock, Settings, Simple Gallery Pro, AndroidWorld, Broccoli, Clipper, MiniWoB, OpenTracks, APKPure, Breezy Weather, Coursera, DuckDuckGo, F-Droid, NerdCalci, Snapseed, Quark, Xiachufang \\
Productivity (23) & Files, Contacts, Dialer, Simple SMS Messenger, Simple Calendar Pro, Joplin, Markor, Tasks, Simple Draw Pro, Audio Recorder, Amaze, Material Files, MiXplorer, MT Manager, Etar, Omni Notes FOSS, Code Editor, DeepL, GeminiAssist, Doubao, Element, Mattermost, Session \\
Entertainment (15) & VLC, Retro Music, iQIYI, Youku, Bilibili, NetEase Cloud Music, NewPipe, AntennaPod, Toutiao, Yahoo News, Yahoo Sports, Wikipedia, Librera, 2048, Angry Birds Friends \\
Shopping (1) & eBay \\
Travel (3) & OsmAnd, Booking.com, Citymapper \\
Finance (1) & Pro Expense \\
\bottomrule
\end{tabular}
\end{table}

\section{Data Collection}
\label{app:data}

\subsection{Construction details}
\label{app:sft-data}

\paragraph{Environment representation and task synthesis.}
Task synthesis is grounded in the functions an application supports and the states reachable from a given screen. For Android emulator applications, exploration is used to map screens and transitions. For \textsc{MobileGym++}, the page, action, and business-object declarations are available from each simulated client. A global view summarizes an application's features and where they are located; a local view lists states directly reachable from the current screen. Instructions are phrased as user goals rather than naming the controls, tools, or verifier fields to use.

\paragraph{Trajectory rollout and filtering.}
The same model executes each instruction step by step, producing a \texttt{Thought:} paragraph followed by one action per turn. When the foreground application exposes app-native tools, its tool catalog is available and the model may choose a tool call or GUI action. Both channels act on the same application state. A rollout is retained when the task succeeds under its runtime-specific check: the programmatic verifier for \textsc{MobileGym++} or a vision-language judge for Android emulator tasks. Degenerate transitions that leave application state unchanged are removed.

\subsection{Executable RL tasks}
\label{app:rl-data}
Each executable RL task is a triple of instruction, initial state, and verifier. The initial state is a snapshot of the \textsc{MobileGym++} application stores, so repeated rollouts of a task start from the same state. The verifier compares application state before and after an episode and checks whether the requested change was written. Since it reads application state rather than the screen, it evaluates GUI and app-native tool actions alike.

\subsection{Data composition}
\label{app:data-quality}

\paragraph{Source coverage.}
The SFT mixture contains 11,643 trajectories. Of these, 7,725 come from emulator(about two third), while 3,918 come from simulator (about one third; Table~\ref{tab:app-sft-data}). The mixture therefore combines trajectories from Android emulator applications with controllable simulated-client workflows.

\paragraph{Trajectory length and action channels.}
Mean trajectory lengths range from 6.8 steps for hybrid trajectories to 11.9 for AndroidWorld, with MobileGym at 10.7 and MobileWorld at 8.8. The 1,774 hybrid trajectories use app-native tools alongside GUI actions.

\section{MobileGym++ Bench}
\label{app:mgpp-bench}

The following subsections document the task split and domain distribution for MobileGym++ Bench, and give two worked examples that illustrate the evaluation protocol concretely.

\subsection{Task composition}
\label{app:mgpp-bench-split}
\begin{figure}[t]
\centering
\includegraphics[width=\linewidth]{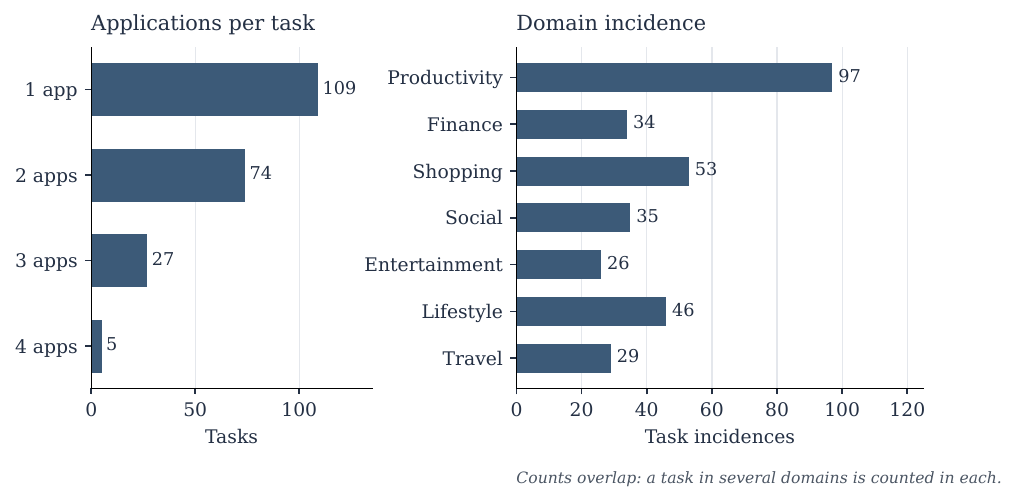}
\caption{MobileGym++ Bench composition. Left: tasks by the number of applications they touch. Right: task incidences by domain. Domain counts are not mutually exclusive.}
\label{fig:bench-composition}
\end{figure}

The 215 tasks span one to four applications. Of these, 109 use one app, 74 use two, 27 use three, and 5 use four; 106 tasks therefore require cross-app interaction. Figure~\ref{fig:bench-composition} and Table~\ref{tab:bench-domain-incidence} report the same split and the task incidences by domain. Because a cross-domain task is counted once in each domain it touches, these incidences overlap and do not sum to 215.

\begin{table}[t]
\centering
\small
\caption{MobileGym++ Bench task incidences by domain. A task that touches more than one domain is counted in each, so the column does not sum to 215.}
\label{tab:bench-domain-incidence}
\begin{tabular}{lr}
\toprule
Domain & Task incidences \\
\midrule
Productivity & 97 \\
Finance & 34 \\
Shopping & 53 \\
Social & 35 \\
Entertainment & 26 \\
Lifestyle & 46 \\
Travel & 29 \\
\bottomrule
\end{tabular}
\end{table}

\subsection{Example tasks}
\label{app:mgpp-bench-ex}

\paragraph{GUI-only: transfer search results into Notes.}
Task 202 (\texttt{HTV6Case27}) asks the agent to search RedNote (Xiaohongshu) using the Chinese term for travel, identify the two most-liked posts among the first ten results, and write their full titles to the system Notes app, one per line (not to flomo). In the GUI-only setting, the agent can use the \texttt{mobile\_use} interface actions but receives no app-native tools. The programmatic checker independently retrieves the top two titles from the initial search results and verifies that both appear in the latest Notes entry.

\paragraph{Hybrid: complete a Taobao purchase.}
Task 111 (\texttt{HTV6Case69}) asks the agent to buy a Jianmu large-capacity insulated flask made of 316 stainless steel, choose the saved address marked School, and pay using Huabei installments. The hybrid action set combines GUI actions with Taobao app-native tools; relevant calls include \texttt{taobao\_\_search\_items}, \texttt{taobao\_\_get\_item\_summary}, and \texttt{taobao\_\_add\_to\_cart}. \texttt{taobao\_\_prepare\_checkout} opens the confirmation page as a non-persisted draft, so the agent must still complete the payment flow in the GUI. The checker requires a new paid order for the target item, with the school-tagged address and Huabei installment payment recorded in the order.

\section{App-native tool examples}
\label{app:tool-schema}
App-native tools occupy the same action slot as \texttt{mobile\_use} GUI interactions: at each step, the agent chooses one action from a single list that includes both. When an application is in the foreground, its tool catalog is appended to that list; when the application leaves the foreground, its tools are removed. The following Taobao schemas illustrate read, state-changing, and navigation-to-GUI operations. The tool handlers and the visible application use the same \texttt{useTaobaoStore}: for example, \texttt{add\_to\_cart} updates the cart state that the GUI renders. Tool schemas reject undeclared input keys.

\begin{lstlisting}[basicstyle=\ttfamily\small,frame=single,breaklines=true]
taobao__search_items(query: string, limit?: integer)
taobao__get_item_summary(item_id: string)
taobao__add_to_cart(item_id: string, sku_description: string,
                    quantity: integer [1..99])
taobao__prepare_checkout(cart_line_ids: string[1..30])
\end{lstlisting}

The first two calls read the catalog; \texttt{add\_to\_cart} validates the item, SKU, and quantity before writing the shared cart. \texttt{prepare\_checkout} accepts one to 30 valid cart-line IDs, opens the confirmation page, and creates neither an order nor a payment. The GUI then continues from the same app state.

\section{System prompt}
\label{app:sys-prompt}

The prompt below is the evaluation harness for Qwen3.5-9B, Qwen3.6-27B, and the OpenMobile-2 checkpoints trained from them. It is used in both the GUI-only and hybrid settings. The GUI tool \texttt{mobile\_use} is always available; when an application exposes app-native tools, the environment lists them in the tool message for as long as that application stays in the foreground. In the GUI-only setting no such catalog is attached. Results marked $*$ for other models in Table~\ref{tab:main} are reproduced by us, but not with this harness. Each of those models keeps its own action format and system prompt.

\promptfile{System prompt for Qwen and OpenMobile-2 evaluation}{prompts/system_prompt.txt}

\end{document}